\documentclass[11pt,nofootinbib,aps,preprint]{revtex4-1}

\usepackage{amsmath,mathrsfs}
\usepackage{graphicx,epsfig}
\usepackage{latexsym,amssymb}
\usepackage{epsf}
\usepackage{ifpdf,natbib,lineno}

\begin{document}

\title{{\bf  Non-Topological  Solitons   in $3+1$ Dimensions}}

\author{ M. Mohammadi$^{1}$} \email{physmohammadi@pgu.ac.ir}   \author{A. R. Olamaei$^{2}$}\email{olamaei@jahromu.ac.ir} \address{$^1$Physics Department, Persian Gulf University, Bushehr 75169, Iran. \\ $^2$Physics Department, Jahrom University, Jahrom, P. O. Box 74137-66171, Iran.}

\begin{abstract}

The paper, classically,  presents a   special  stable non-topological solitary wave packet solution  in $3+1$ dimensions for an extended  complex non-linear  Klein-Gordon (CNKG) field system.   The rest energy of this special solution is minimum among  other (close) solutions i.e. it is a soliton solution. The equation of motion and other properties for this special stable solution are reduced  to the same original known  CNKG system.


\end{abstract}

\maketitle


\section{Introduction}

In soliton paradigm the relativistic classical field theory is an attempt to model particles in terms of non-singular, localized solutions of properly tailored non-linear PDEs \cite{rajarama,Das,lamb,Drazin}.   Classically, a particle is considered as a rigid body which obeys  the famous relativistic energy-momentum relation  and survives  in elastic  collisions.  To find a soliton solution,  like classical particles, firstly, we  try to find a solitary wave solution with a localized energy-density; secondly, it must be checked out whether it is  stable or not.

 A solitary wave solution is  stable if  the related  rest  energy is minimized against any arbitrary small permissible   deformation.
As an example of this definition, kink and anti-kink solutions of  the real non-linear Klein-Gordon  systems in 1+1 dimensions are  stable objects. It was shown theoretically and numerically for  such  systems that,  the solitary kink and anti-kink solutions are  stable objects \cite{Riazi,GH,MM1,MM2}. In 3+1 dimensions, unfortunately, only a few models of the known non-linear PDEs  with soliton solutions have been proposed, among which, one can mention the Skyrme model of baryons \cite{SKrme,SKrme2} and 't Hooft Polyakov model which yields magnetic monopole  solitons \cite{toft,pol}.

In this paper we  reintroduce  the  complex nonlinear Klein-Gordon (CNKG) systems with non-topological solitary wave-packet solutions (SWPS's) \cite{Lee0,Lee,Lee2,W,W2,Scoleman,Lee3,Riazi2,Ab,MM3}. For any CNKG system, there are infinite types of SWPS which can be identified by different rest frequencies $\omega_{o}$  and electrical charges. However, it  has   not been introduced   yet a CNKG model with a stable SWPS.  Some of references called  these solutions Q-balls or Q-solitons \cite{Scoleman}. Although, it was shown that the SWPSs or Q-balls   have  the minimum rest energies  among the other solutions with the same electrical charge, it is not a sufficient condition to show that Q-ball solutions are stable objects, as they are not stable under any arbitrary small deformation.  They can emit some small localized perturbations (for example in a collision processes) which without the violation of energy and   electrical charge conservation, turn to other solutions  with less rest energies and electrical charges, i.e.  they are not essentially stable objects.

For simplicity, we will study  a special  CNKG system with many Gaussian  SWPSs which can be identified with different rest frequencies $\omega_{o}$. It  will be shown  that for such solutions, essentially there is no  stable SWPS, i.e. it  is not possible to find a special SWPS (SSWPS) for which its rest energy being minimum among the other close solutions. All of the close  solutions of a  SSWPS are  permissible small  deformations (variations) of that, which are again solutions themselves.

In this paper,  it will be  shown that  to have a stable SSWPS with a standard known  CNKG equation of motion as the dominant dynamical equation, we have to consider the original  CNKG Lagrangian density with  three  additional terms (an extended CNKG system). These additional terms alone (i.e. without the original CNKG Lagrangian density) lead  to a zero rest mass soliton solution  which can move at any arbitrary speed (not greater  than the light speed). In other words, for the new extended CNKG system, these additional terms behave  like a  zero rest mass spook\footnote{We chose  the word "\emph{spook}" in order to not to confuse with words like  "\emph{ghost}" and "\emph{phantom}", which have meaning in the  literature}  which surrounds the SSWPS and resists to any arbitrary deformation. In fact, for the new extended system, there are new complicated equations of motion with different solutions, but for one of them (i.e. for the SSWPS), the equations of motion and all of the other properties  would be reduced  to the same original  ones (i.e. the same  CNKG  equations and properties). In this  new model,
there are three   parameters $A_{i}$'s ($i=1,2,3$) which larger values of those   lead  to   more stability of the SSWPS, i.e.  the difference between the rest energy of the SSWPS and  the rest energies  of the other close solutions increases  with increasing the amount of   $A_{i}$'s.

The organization of this paper is as follows: In the next section,  for  the CNKG systems  we will set up the basic equations and consider general properties of the related solitary wave-packet solutions. In section III,  we  have some arguments about stability concept. In section IV,  an extended CNKG system will be introduced to obtain a stable SSWPS for which the dominant  equations of  motion are reduced  to the same original   CNKG versions. In section V, the stability of the SSWPS will be considered specially  for small deformations.  The last section is devoted to  summary and conclusions.

\section{ Basic properties of  the CNKG systems}\label{sec2}

The present calculations are based on a relativistically  U(1)-Lagrangian density  in $3+1$ dimensions:
 \begin{equation} \label{lag}
{\cal L}= \partial_\mu \phi^*
\partial^\mu \phi -V(|\phi |) ,
 \end{equation}
in which $\phi$ is a complex scalar field and $V(|\phi |)$, the field potential, is a self-interaction term  which depends only on the modulus of the scalar field. By varying this action with respect to $\phi^{*}$, one obtains
the field equation
\begin{equation} \label{eq}
 \Box \phi =\frac{\partial^2\phi}{\partial
  t^2}- \nabla^{2} \phi=-\frac{\partial V}{\partial
  \phi^*}=-\frac{1}{2}\frac{d V}{d|\phi|}\frac{\phi}{|\phi|},
  \end{equation}
which is the complex non-linear  Klein-Gordon equation in $3+1$ dimensions. Note that, through the paper,   we take the speed of light equals to one.  To simplify  Eq.~(\ref{eq}), we can change variables to the polar fields $R(x^{\mu})$ and $\theta(x^{\mu})$ as defined by
\begin{equation} \label{polar}
 \phi(x,y,z,t)= R(x,y,z,t)\exp[i\theta(x,y,z,t)].
\end{equation}
In terms of  polar fields, the Lagrangian-density and related field equations are reduced  respectively to
\begin{equation} \label{Lag2}
{\cal L}=(\partial^\mu R\partial_\mu R) +R^{2}(\partial^\mu\theta\partial_\mu\theta)-V(R),
\end{equation}
and
\begin{eqnarray} \label{e25}
&&\Box R-R(\partial^\mu\theta\partial_\mu\theta)=-\frac{1}{2}\frac{dV}{dR}, \\&&
\partial_{\mu}(R^2\partial^{\mu}\theta)=2R(\partial_{\mu}R\partial^{\mu}\theta)+R^{2}(\partial^\mu\partial_\mu\theta)=0.
\end{eqnarray}
The related Hamiltonian (energy) density  is obtained via the Noether's theorem:
\begin{eqnarray} \label{TE}
&&T^{00}=\varepsilon(x,t)=\dot{\phi}\dot{\phi}^{*}+\nabla\phi \cdot \nabla\phi^{*}+V(|\phi |) \nonumber\\&&
=(\dot{R}^2+\nabla R \cdot \nabla R)+R^2(\dot{\theta}^2+\nabla\theta \cdot \nabla\theta)+V(R),
\end{eqnarray}
where dot  denotes differentiation with respect to $t$.

We would like to consider systems with spherically symmetric solitary wave-packet solutions i.e. the ones for which the related modulus and phase  functions, when they are  at rest,  are represented as follows:
\begin{equation} \label{So1}
R(x,y,z,t)=R(r)=R(\sqrt{x^2+y^2+z^2}),\quad  \quad\theta(x,y,z,t)=\omega_{o}t,
\end{equation}
in which $R(r)$ must be  a localized function. For  anstaz (\ref{So1}), the related equation of motion (6) is satisfied automatically and would  reduced  to
 \begin{equation} \label{Re}
 \dfrac{1}{r^2}\dfrac{d}{dr}(r^2\dfrac{dR}{dr})=\frac{1}{2}\frac{dV}{dR}-\omega_{o}^{2}R.
 \end{equation}
 Depending on  different values of   $\omega_{o}$, different solutions  for $R(r)$ can be obtained. Accordingly,  there are a continuous range of different  solitary wave packet solutions  with different rest frequencies ($\omega_{o}$).
A moving solitary wave packet solution can be obtained easily by a relativistic boost. For example, a solitary wave packet solution, with rest frequency $\omega_{o}$, which moves in the x-direction with a constant velocity $\textbf{v}=v\widehat{i}$, would take the form:
\begin{equation} \label{So}
R(x,y,z,t)=R(\sqrt{\gamma^2(x-vt)^2+y^2+z^2}),\quad  \quad\theta(x,y,z,t)=k_{\mu}x^{\mu},
\end{equation}
in which  $\gamma=1/\sqrt{1-v^2}$, and  $k^{\mu}\equiv(\omega,\textbf{k})=(\omega,k,0,0)$ is a $3+1$ vector, provided
\begin{equation} \label{pro}
\textbf{k}=k\widehat{i}= {\omega}\textbf{v},
\end{equation}
and
\begin{equation} \label{pro2}
\omega=\gamma\omega_{o}.
\end{equation}

\begin{figure}[ht!]
  \centering
  \includegraphics[width=110mm]{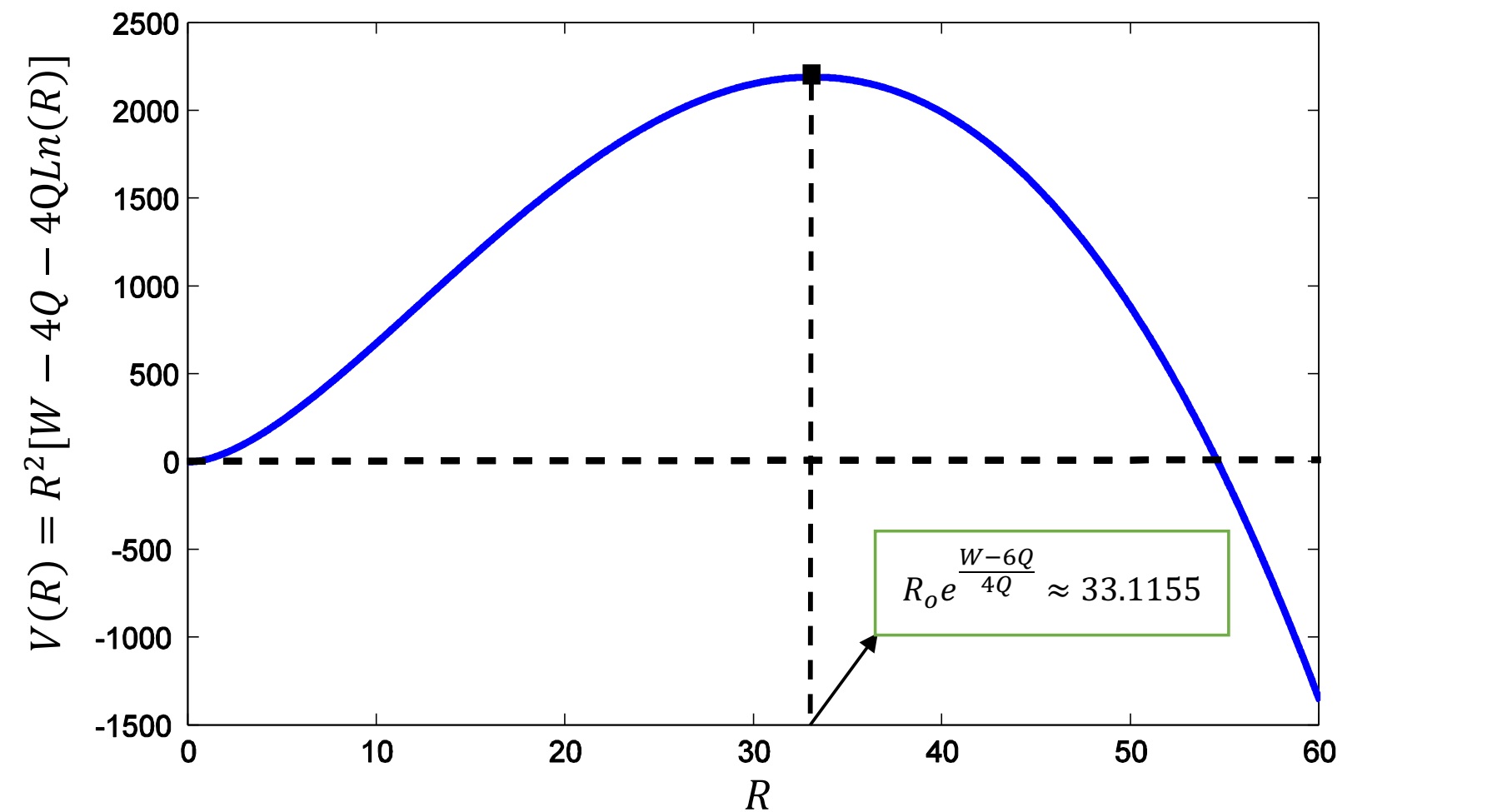}
  \caption{The field potential (\ref{fp}) versus $R$. we have set $W=20$, $Q=1$ and $R_{o}=1$. } \label{po}
\end{figure}

For simplicity,  to obtain some  generic Gaussian function   as  different proposed  solitary wave solutions,  one can use the following field potential as an example of the nonlinear KG (nKG) self interacting fields:
\begin{equation} \label{fp}
V(R)= R^{2} \left[ W-4Q-4Q\ln(\frac{R}{R_{o}})\right],
\end{equation}
in which, $W$, $R_{o}$ and $Q$ are  some arbitrary constants. By solving  equation (\ref{Re}),   the variety of    solitary wave packet solutions as a function of  $\omega_{o}$ can be obtained:
\begin{equation} \label{f}
R(r)=\sigma(\omega_{o}) R_{o}e^{-Qr^{2}},
\end{equation}
where $\sigma(\omega_{o})= \exp(\frac{W-\omega_{o}^{2}}{4Q})$. Note that, any arbitrary   solitary wave solution is not necessarily    stable. For example if we set $W=20$, $Q=1$ and $R_{o}=1$, the related  potential (\ref{fp}) as a function of  $R$ is shown in Fig.~\ref{po}, which its maxima   occur at $R=R_{o}\exp (\frac{W-6Q}{4Q})\approx 33.1155$
(c.f. figure \ref{po}). According to  general theory of  the classical relativistic field theory, it is easy to conclude that the solutions for which $R_{\textrm{max}}>33.1155$, are not essentially stable and can evolve to infinity without the violation of the energy conservation law.
To be specified, among infinite types of  solitary wave-packet solutions (\ref{f}),  a special one, $\omega_{o}=\omega_{s}$, with $\omega_{s}^2=W=20$ will be considered. For this special solution $R_{\textrm{max}}=1$ which is clearly less than $33.1155$. Therefore, for the special solitary wave-packet solution ($\omega_{s}=\pm\sqrt{20}$) and those  solutions which are close to that, we are sure that the possible values  of $R$ are in  the   domain of   potential (\ref{fp}),  which   is increasingly positive, that  is  a necessary  condition for the stability of the  special solitary wave-packet solution (SSWPS).

   The travelling solitary wave packet solution with rest frequency $\omega_{o}$, is obtained by applying a Lorentz boost. For motion in the x-direction, we obtain
\begin{equation} \label{SS}
\phi(x,y,z,t)=\sigma R_{o}e^{(-Q[\gamma^2(x-vt)^2+y^2+z^2])} \exp{(i\omega_{o}\gamma[t-vx])}=
R(r)e^{(i\omega_{o}\widetilde{t}~)},
\end{equation}
in which $\widetilde{t}=\gamma[t-vx]$. The total energy of a non-moving  solitary wave packet solution can be obtained and equated to the rest energy  of the related
particle-like configuration  as
\begin{eqnarray} \label{fe}
&&E_{o}=m_{o}c^{2} \equiv \int T^{00} d^{3}\textbf{x}=\int \left[(\nabla R \cdot \nabla R)+R^2(\dot{\theta}^2)+V(R)  \right]d^{3}\textbf{x}\nonumber \\&&
\quad\quad\quad=\int_{0}^{\infty} \left[(\frac{dR}{dr})^{2}+\frac{\omega_{o}^{2}}{c^{2}}R^{2}+V(R)\right]4\pi r^{2}dr \nonumber \\&&\quad\quad\quad=\int_{0}^{\infty} \left[2R_{o}^{2}\sigma^{2}e^{(-2Q r^{2})}\left(4Q^{2}r^{2}-2Q+\omega_{o}^2\right)\right]4\pi r^{2}dr \nonumber \\&&\quad\quad\quad
=2R_{o}^{2}\sigma^{2}\left(\frac{\pi}{2Q}\right)^{\frac{3}{2}}(Q+\frac{\omega_{o}^{2}}{c^{2}}).
\end{eqnarray}
Generally, it is possible to show that   for each localized solitary wave solution in the relativistic classical field theory,  which the related energy-momentum tensor $T^{\mu\nu}$ asymptomatically approaches to zero at  infinity,  four independent integrations of the energy-momentum tensor components $T^{\mu 0}$ over  the whole  space,  form  components of  a four vector. Therefore, generally we expect the following relations to be satisfied for a moving solitary wave packet solution:
\begin{eqnarray} \label{d}
&&E=m c^{2}\equiv \int T^{00} d^{3}\textbf{x}=\gamma E_{o}=\gamma m_{o}c^{2}, \\&&
\textbf{p}\equiv \int \left(T^{01},T^{02},T^{03} \right)d^{3}\textbf{x}=\gamma m_{o} \textbf{v}.
\end{eqnarray}

It is worth to mention that  equations (\ref{pro2}) and  (\ref{d}) show that the energy and  the frequency are possessing the same behavior and we can relate them via introducing a Planck-like constant $\overline{h}$:
\begin{equation} \label{dc}
 E=\overline{h} \omega.
\end{equation}
It is easy to understand that $\overline{h}$ is a function of  rest frequency $\omega_{o}$ and for different solitary  wave-packet solutions, there are different $\overline{h}$ constants. Similarly, it is possible to find a relation between relativistic momentum of a solitary wave packet  solution and the wave number $\textbf{k}$:
\begin{equation} \label{md}
\textbf{p}=\overline{h} \textbf{k}.
\end{equation}
This equation is  interesting, since it resembles the deBroglie's relation.

The Lagrangian density in Eq.  (1) is $U(1)$ invariant like electromagnetic theory  and this yields to conservation of electrical  charge. So, according to the Noether theorem, we can introduce a conserved electrical   current density as
\begin{equation} \label{cur}
j^\mu\equiv i\eta (\phi^*\partial^\mu \phi-\phi\partial^\mu \phi^*)=-2\eta (R^{2}\partial^\mu \theta),\quad  \partial_\mu j^\mu=0,
\end{equation}
in which $\eta$ is a constant included for dimensional reasons \cite{Riazi2}, and  the corresponding conserved charge would be
\begin{equation} \label{Bar}
q=\int_{-\infty}^{+\infty}j^0 d^{3}\textbf{x}=\int_{-\infty}^{+\infty}i\eta(\phi^*\dot{\phi}-\phi\dot{\phi}^*)d^{3}\textbf{x}.
\end{equation}
 It is notable that  both  positive and negative signs of $|\omega_{o}|$  (i.e. $\omega_{o}=\pm |\omega_{o}|$) lead to the same  solution for the differential equation (\ref{Re}). They have the same rest mass (energy) but different electrical   charges (positive and negative). It is easy to show that for the solutions with $\omega_{o}>0$ ($\omega_{o}<0$), if we take $\eta>0$,  electrical  charge is negative (positive). This shows that the positive and negative solutions are particle and anti-particle.

\section{   stability considerations}\label{sec3}


In  soliton paradigm,  particles are considered as localized solutions of the relativistic nonlinear field equations. In this paradigm, the main goal is  to find a stable solitary wave solution  or a soliton solution.  We can define a solitary wave solution of a relativistic nonlinear field equation as \emph{stable}, if its rest energy being minimum among the other close solutions. The close solutions, of a special solitary wave solution  which are all permissible small deformations (variations) of that, are again solutions of the equations of motion. For example, if one consider the previous   CNKG systems with standard equations of motion (5) and (6), the  close solutions $\phi=Re^{i\theta}=\phi_{s}+\delta\phi=(R_{s}+\delta R)e^{i(\theta_{s}+\delta\theta)}$ of a special solution $\phi_{s}=R_{s}e^{i\theta_{s}}$, are ones for which we have
\begin{eqnarray} \label{hij}
&&\Box (R_{s}+\delta R)-(R_{s}+\delta R)(\partial^\mu(\theta_{s}+\delta\theta)\partial_\mu(\theta_{s}+\delta\theta))=-\frac{1}{2}\frac{dV(R_{s}+\delta R)}{d(R_{s}+\delta R)}, \\&&
\partial_{\mu}((R_{s}+\delta R)^2\partial^{\mu}(\theta_{s}+\delta\theta))=0,
\end{eqnarray}
where $\delta R$ and $\delta \theta$ can be any  permissible   space-time variations  which satisfy these equations simultaneously.  To first order of  variations, they yield
\begin{eqnarray} \label{sdfg}
&&\Box (\delta R)-(\delta R)(\partial^\mu\theta_{s}\partial_\mu\theta_{s})-2R_{s}(\partial^\mu\theta_{s}\partial_\mu(\delta\theta))\approx-\frac{1}{2}\frac{d^2V(R_{s})}{dR_{s}^2}(\delta R), \\&&
\partial_{\mu}(R_{s}^2\partial^{\mu}(\delta\theta)+2R_{s}\delta R\partial^{\mu}\theta_{s})\approx 0,
\end{eqnarray}
which is the right PDE's  to specify the  small permissible deformations (close solutions).

Based on this definition,  for  different solitary wave packet solutions (\ref{f}), if one plot rest energy $E_{o}$ versus $\omega_{o}^2$ (\ref{fe}),
 the resulted  curve (Fig.~\ref{TE}) shows that there is not a trivial solitary wave packet solution with a  minimum rest energy. For the case $\omega_{o}=0$, we can see a minimum, but for this special case we encounter with a real system which  Virial  theorem \cite{DRe}  essentially prevents us from having a stable solution. In fact, for the case $\omega_{o}=0$, the maximum value of modulus  function (\ref{SS}) is  $R_{\textrm{max}}= R_{o}\exp(\frac{W}{4Q})$, that is greater  than  $R_{o}\exp (\frac{W-6Q}{4Q})$,  means that it is not a stable solitary wave solution.
\begin{figure}[ht!]
  \centering
  \includegraphics[width=100mm]{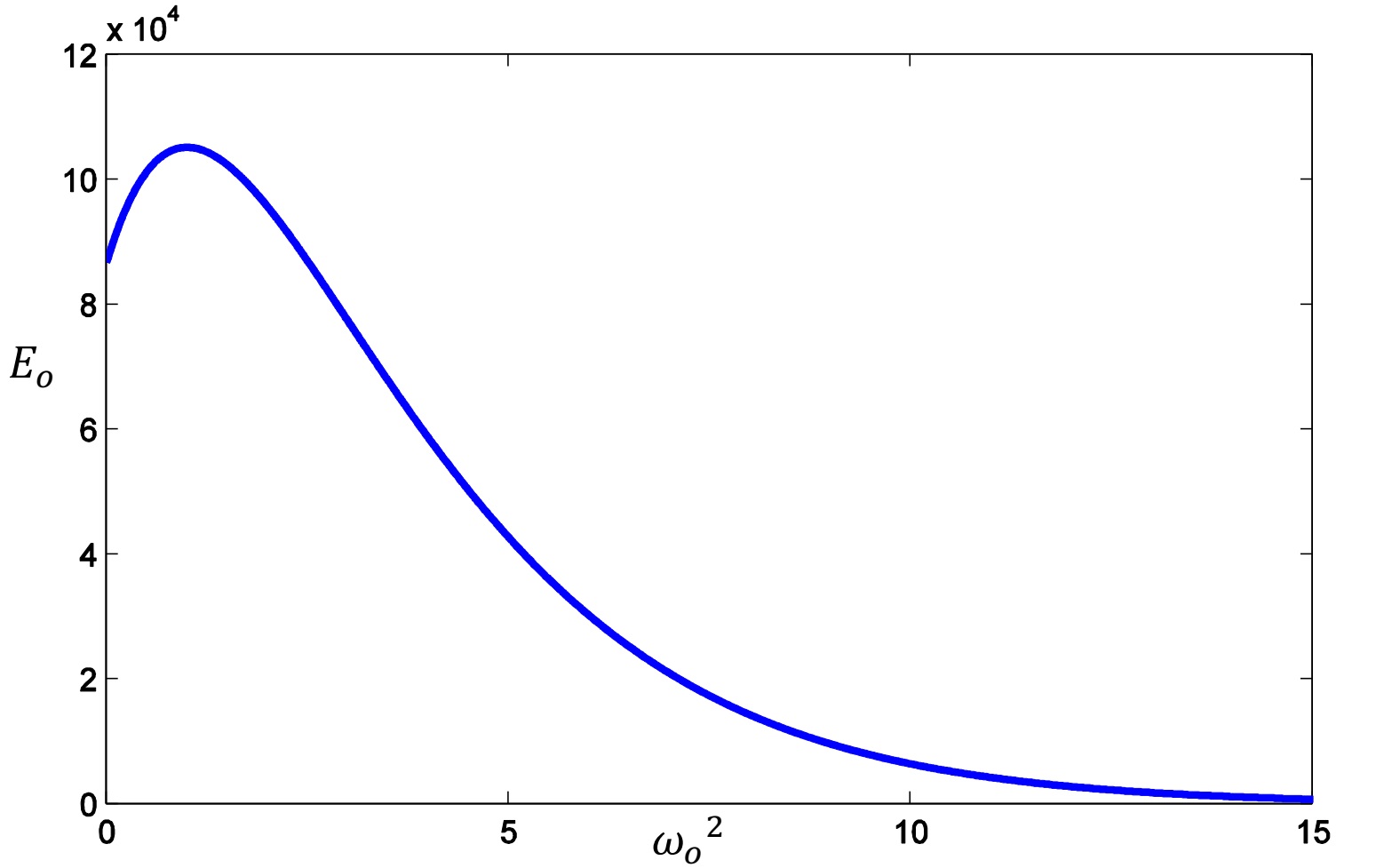}
  \caption{Rest energy $E_{o}$ versus $\omega_{o}^2$, with $W=20$, $Q=1$ and $R_{o}=1$.} \label{TEs}
\end{figure}

 In soliton paradigm to overcome  the  stability  problem, usually   the topological solitary wave solutions have been searched.
For example,  the famous kink (anti-kink) solutions of the real non-linear KG systems, Skyrme model and 't Hooft-Polyakov model are few examples which finally yield to topological stable solitary wave solutions.  The non-topological solutions are more interesting, because a many particle-like  solution  can be easily
constructed just by adding many far enough distinct  solitary wave solutions together. There are usually  hard and complicated conditions for topological solitons to provide a many particle-like  solutions. However, if one considers a non-topological solitary wave solution of a  nonlinear relativistic field system like  a fundamental particle,  its rest energy must be minimum among other (close) solutions. In this paper, mathematically,  we will introduce a nonlinear  relativistic field system which leads to a (non-topological)  special   stable solitary wave-packet solution. Moveover, since many of  the dynamical equations of the  particles in quantum field theory are KG  or nonlinear KG (-like), we expect  that the right dominant dynamical  PDE for the free special solitary wave packet solution  (SSWPS)  have to remain  the same as CNKG equation of motion (\ref{eq}) as well.

\section{A new extended CNKG system  with a special   stable solitary wave packet solution}\label{sec4}

 There are many  known particles in the nature. Standard model (SM) is the  successful theory which is used to describe these particles. For every   type of known particles, there is just a specific   field equation with some specific constants.   For example, the well-known  non-linear  $\phi^4$ theory is used just for Higss particles and  Dirac's equation is used  for electrons and positrons with some inputs (electrons mass and charge). Dirac's equation for other particles like muons and neutrinos  were used again but with different inputs.
 The proper constants for any type of fundamental particles are usually determined in the laboratory and introduced in relevant equation and inputs.

In this paper, motivated by the classical relativistic field theory, we go through the similar procedure. We will show that what kind of constraints are needed for a special type of localized solutions (\ref{f}), to be the only valid stable particle solution for a standard CNKG system (1). As emphasized, the dynamical equation (\ref{eq}), or equivalently Eqs. (5) and (6), have infinite localized solutions (see Eq. (\ref{f})) and none of them are essentially stable. However, our assumption is that, in a new classical field theory, necessary conditions can be came together in such a way that just for one of the infinite wave packet functions (\ref{f}), as a single sloliton solution, the dominant dynamical equation to be  of the standard form (\ref{eq}), and the rest of them would not be the solution of the new system anymore. For more clarification, assume that in a way (like observing a new particle in a Lab), a particle-like answer could be introduced as a stable localized solution of an unknown classical field system as follows:
\begin{equation} \label{f2}
\phi_{s}(x,y,z,t)=R_{s}(r)e^{(i\theta_{s})}=
R_{s}(r)e^{(i\omega_{s}\widetilde{t}~)}=\sigma_{s} R_{o}e^{(-Q r^{2})} e^{(i (k_{s})_{\mu}x^{\mu})}=e^{-r^{2}} e^{(i (k_{s})_{\mu}x^{\mu})},
\end{equation}
where $r=\sqrt{\gamma^2(x-vt)^2+y^2+z^2}$ (when it moves  in the $x$-direction). For simplicity, we set $W=\omega_{s}^2=20$ and $Q=R_{o}=1$, then $\sigma_{s}= \exp(\frac{W-\omega_{s}^{2}}{4Q})=1$. In fact, it is a special solution from the  infinite solutions  (\ref{f}) of the dynamical equation (\ref{eq}), which the  subscript $s$ indicating the "\emph{special}" to emphasis on this point. Here we consider three assumptions: First, we assume that in general, the Lagrangian density of the relativistic field has a new and complex form (${\cal L}_{N}$), different from that of  Eq.~(\ref{Lag2}), with the condition that the special wave packet (\ref{f2}) still be one of its valid solutions.  Second, we assume that the new complicated Lagrangian density and its resulting dynamical equations, just for the special solution (\ref{f2}), would be reduced to the standard form of Eqs. (\ref{Lag2}), (5) and (6), respectively. In other words, we expect that the dominant dynamical equations take the form of the standard and well-known Eqs. (5) and (6), only for the special solution (\ref{f2}). Third, we expect that the special solution (\ref{f2}) to be a stable and soliton solution which means that its rest energy is the minimum among other nontrivial solutions of this new system, which means any change in its internal structure would cause the increase in its rest energy. Satisfying these three assumptions, we get just one single stable particle-like solution that its dominant dynamical equations, like many known particles, are in the standard CNKG form.

According to the pervious demands, in general, the new complex form of the lagrangian density (${\cal L}_{N}$) can be considered as the same  original lagrangian density  (\ref{Lag2}) plus   a new unknown   functional term $F$:
\begin{equation} \label{LN}
 {\cal L}_{N}= {\cal L}+F=\left[\partial^\mu R\partial_\mu R +R^{2}(\partial^\mu\theta\partial_\mu\theta)-V(R)\right]+ F.
 \end{equation}
In other words,  we are going to find an additional  proper term $F$ for  the original  lagrangian density (\ref{lag}), with a  special potential (\ref{fp}),  for which pervious demands satisfied generally.
 The unknown scaler $F$ must be, in general,  function of  all possible allowed scalar structures. The  scalars which can appear in the Lagrangian density are  modulus field $R$, phase field $\theta$,  $\partial_{\mu}R\partial^{\mu}R$, $\partial_{\mu}\theta\partial^{\mu}\theta$ and $\partial_{\mu}R\partial^{\mu}\theta$. As we indicated  before,  we suppose that  one of the pervious solitary wave packet solutions (\ref{f}), with a special rest frequency $\omega_{o}=\omega_{s}$ (\ref{f2}), to be a solution again and all relations and equations which were derived in  section \ref{sec2} for this SSWPS (\ref{f2}) would stay unchanged. Therefore, at the first step, for the  SSWPS (\ref{f2}), since we expect  the new Lagrangian density (\ref{LN}) to be reduced to  the  primary original version (\ref{Lag2}), we conclude that the additional term $F$ must be zero for the SSWPS  (\ref{f2}). The new equations of motion for the new Lagrangian density (\ref{LN}) are
\begin{eqnarray} \label{geq}
&&\Box R-R(\partial^\mu\theta\partial_\mu\theta)+\frac{1}{2}\frac{dV}{dR}+\frac{1}{2}\left[ \frac{\partial}{\partial x^{\mu}}\left(\frac{\partial F }{\partial (\partial_{\mu}R)}\right)-\left(\frac{\partial F}{\partial R}\right)  \right]=0,\\&&  \partial_{\mu}(R^2\partial^{\mu}\theta)+\frac{1}{2} \left[ \frac{\partial}{\partial x^{\mu}}\left(\frac{\partial F }{\partial (\partial_{\mu}\theta)}\right)-\left(\frac{\partial F}{\partial \theta}\right)  \right]=0.
\end{eqnarray}
To be sure that the  conservation of the electrical charge still remains  valid  in the new extended system (\ref{LN}), the functional $F$ must not dependent on $\theta$. Therefore, the new electrical current in the new extended system (\ref{LN}) is
\begin{eqnarray} \label{cure}
j^{\mu}=R^2\partial^{\mu}\theta+\frac{1}{2}\left[ \frac{\partial}{\partial x^{\mu}}\left(\frac{\partial F }{\partial (\partial_{\mu}\theta)}\right)  \right].
\end{eqnarray}
Moreover, to be sure that the SSWPS (\ref{f2}) is  a solution again,  we expect for the SSWPS  (\ref{f2}) all additional terms  which appeared in the square brackets $[\cdots]$ become zero. Since the functional $F$ is considered essentially  different  from the original Lagrangian density (\ref{lag}), i.e. $F$ is not linearly dependent on ${\cal L}$,   we conclude that  all distinct terms $\frac{\partial}{\partial x^{\mu}}\left(\frac{\partial F }{\partial (\partial_{\mu}R)}\right)$, $\frac{\partial F}{\partial R}$ and $\frac{\partial}{\partial x^{\mu}}\left(\frac{\partial F }{\partial (\partial_{\mu}\theta)}\right)$ must be  zero independently  when we have the SSWPS  (\ref{f2}). It means that   the dominant  equations of motion, for the SSWPS (\ref{f2}),  would be the same standard CNKG equations of motions (5) and (6), as we expected.   Therefore, $F$ and its  derivatives which appeared  in the pervious equations must be zero for the SSWPS (\ref{f2}). For all these constraints to be satisfied, we have to take  $F$ as a function of powers of ${\cal S}_{i}$'s  (${\cal S}_{i}^{n}$'s with $n\geq3$), where    ${\cal S}_{i}$'s ($i=1,2,3$) are three special independent  scalars
\begin{eqnarray} \label{sc2}
 &&{\cal S}_{1}=\partial_{\mu}\theta\partial^{\mu}\theta-20,     \\&&
{\cal S}_{2}=\partial_{\mu}R\partial^{\mu}R-4R^{2}\ln(R), \\&&
{\cal S}_{3}=\partial_{\mu}R\partial^{\mu}\theta,
\end{eqnarray}
which for the SSWPS (\ref{f2}) would be zero. It is straightforward  to show that these special scalars all are equal to zero for the SSWPS (\ref{f2}). For simplicity, if one  considers  $F$ as a function of  arbitrary n'th power of  ${\cal S}_{i}$'s, i.e. $F=F({\cal S}_{1}^{n},{\cal S}_{2}^{n},{\cal S}_{3}^{n})$, it yields
\begin{eqnarray} \label{sc3}
 && \frac{\partial}{\partial x^{\mu}}\left(\frac{\partial F }{\partial (\partial_{\mu}R)}\right)=\sum_{i=1}^{3} \left[n(n-1){\cal S}_{i}^{(n-2)}\frac{\partial {\cal S}_{i}}{\partial x^{\mu}}\frac{\partial {\cal S}_{i} }{\partial (\partial_{\mu}R)} \frac{\partial F}{\partial Z_{i}}+n{\cal S}_{i}^{(n-1)}\frac{\partial}{\partial x^{\mu}}\left(\frac{\partial {\cal S}_{i} }{\partial (\partial_{\mu}R)} \frac{\partial F}{\partial Z_{i}}\right) \right] \nonumber  \\&&
\frac{\partial F}{\partial R}=\sum_{i=1}^{3}\left[n{\cal S}_{i}^{(n-1)}\frac{\partial {\cal S}_{i}}{\partial R}\frac{\partial F}{\partial Z_{i}}\right]\nonumber\\&&
\frac{\partial}{\partial x^{\mu}}\left(\frac{\partial F }{\partial (\partial_{\mu}\theta)}\right)=\sum_{i=1}^{3} \left[n(n-1){\cal S}_{i}^{(n-2)}\frac{\partial {\cal S}_{i}}{\partial x^{\mu}}\frac{\partial {\cal S}_{i} }{\partial (\partial_{\mu}\theta)} \frac{\partial F}{\partial Z_{i}}+n{\cal S}_{i}^{(n-1)}\frac{\partial}{\partial x^{\mu}}\left(\frac{\partial {\cal S}_{i} }{\partial (\partial_{\mu}\theta)} \frac{\partial F}{\partial Z_{i}}\right) \right] \nonumber.
\end{eqnarray}
where $Z_{i}={\cal S}_{i}^n$. It is easy to understand for $n\geq3$ that all these relations would be zero for the SSWPS (\ref{f2}) as we expected. Accordingly, one can show that   the   general form of the functional  $F$ which satisfies  all needed  constraints, can be introduced by a  series:
\begin{eqnarray} \label{sf0}
 F=\sum_{n_{3}=0}^{\infty}\sum_{n_{2}=0}^{\infty}\sum_{n_{1}=0}^{\infty} a({n_{1},n_{2},n_{3}}){\cal S}_{1}^{n_{1}}{\cal S}_{2}^{n_{2}}{\cal S}_{3}^{n_{3}},
\end{eqnarray}
provided $(n_{1}+n_{2}+n_{3})\geq3$.    Note that, in general the coefficients $a({n_{1},n_{2},n_{3}})$, are arbitrary well-defined functional scalers, i.e. they can be again functions of all possible scalars   $R$, $\partial_{\mu}R\partial^{\mu}R$, $\partial_{\mu}\theta\partial^{\mu}\theta$ and $\partial_{\mu}R\partial^{\mu}\theta$ (except $\theta$ itself).

The  stability  conditions  impose  serious  constraints on  function $F$  which causes to reduce series (\ref{sf0}) to some  special  formats. However, again there are many choices which can  lead to a stable SSWPS (\ref{f2}). Among them,  a simple choice which clearly guarantees the stability of the   SSWPS can be introduced as follows:
 \begin{equation} \label{F}
 F=\sum_{i=1}^{3} A_{i}(
 {\cal K}_{i})^3,
 \end{equation}
where $A_{i}$'s ($i=1,2,3$) are just real positive constants for dimensional reasons and ${\cal K}_{i}$'s are three linear independent combination  of ${\cal S}_{i}$'s:
 \begin{eqnarray} \label{e5}
  &&{\cal K}_{1}=\alpha^2{\cal S}_{1},\\&&
{\cal K}_{2}=\alpha^2{\cal S}_{1}+{\cal S}_{2}, \\&&
{\cal K}_{3}=\alpha^2{\cal S}_{1}+{\cal S}_{2}+2\alpha {\cal S}_{3},
\end{eqnarray}
where $\alpha$ is a  real constant included for dimensional  reasons, but for simplicity we can take it equal to one ($\alpha=1$).  It is  obvious that ${\cal K}_{1}$, ${\cal K}_{2}$ and ${\cal K}_{3}$  are all zero just for the SSWPS (\ref{f2}) with rest frequency $\omega_{s}=\sqrt{20}$.

The energy-density that belongs of  new modified Lagrangian-density (\ref{LN}), for the special choice of the additional function $F$ (\ref{F}),   would be
\begin{eqnarray} \label{MTE}
&&\varepsilon(x,y,z,t)=\left[(\dot{R}^2+\nabla R \cdot \nabla R)+R^2(\dot{\theta}^2+\nabla\theta \cdot \nabla\theta)+V(R)\right]+\nonumber\\&&\quad\quad \quad \quad\sum_{i=1}^{3}\left[3A_{i}C_{i}
{\cal K}_{i}^{2}-A_{i}{\cal K}_{i}^3\right]=\varepsilon_{o}+\varepsilon_{1}+\varepsilon_{2}+\varepsilon_{3},
\end{eqnarray}
which divided  into four distinct  parts and
\begin{equation}\label{cof}
C_{i}=\dfrac{\partial{\cal K}_{i}}{\partial \dot{\theta}}\dot{\theta}+\dfrac{\partial{\cal K}_{i}}{\partial \dot{R}}\dot{R}=
\begin{cases}
\quad\quad 2\dot{\theta}^{2} & \text{i=1}
\\
2(\dot{R}^{2}+\dot{\theta}^2) & \text{i=2}
\\
2(\dot{R}+\dot{\theta})^{2}
 & \text{i=3}.
\end{cases}
 \end{equation}
  After a straightforward calculation, one can obtain:
 \begin{eqnarray} \label{eis}
&&\varepsilon_{1}=A_{1}{\cal K}_{1}^2[5\dot{\theta}^2+(\nabla \theta)^2+20],\\&&
\varepsilon_{2}=A_{2}{\cal K}_{2}^2[5\dot{\theta}^2+5\dot{R}^2+(\nabla \theta)^2+(\nabla R)^2+20+4 R^{2} \ln(R)], \\&&
\varepsilon_{3}=A_{3}{\cal K}_{3}^2[5(\dot{\theta}+\dot{R})^2+(\nabla \theta+\nabla R)^2+20+4 R^{2}\ln(R)].
\end{eqnarray}
Note that, the function $[20+4 R^{2} \ln(R)]$  and other terms in the above equations  all  are non-zero and positive definite. Hence, we conclude that   $\varepsilon_{1}$, $\varepsilon_{2}$ and $\varepsilon_{3}$  are  bounded from below, and the minimum values of them are zero, due to   ${\cal K}_{i}=0$ ($i=1,2,3$).  We will show that just for the nontrivial SSWPS (\ref{f2}) (and the trivial solution $R=0$), $\varepsilon_{i}$'s ($i=1,2,3$)  are zero simultaneously, i.e. for other unknown  nontrivial solutions of the new system (\ref{LN}) at least one of the ${\cal K}_{i}$'s or $\varepsilon_{i}$'s ($i=1,2,3$) would be a nonzero function.
 Again, it is obvious that the related  energy density function (\ref{MTE}) is reduced   to  the same original version (\ref{TE}) as well.

If constants $A_{i}$'s in Eq.~(\ref{F}) are considered to be large numbers, the stability of the SSWPS would be satisfied appreciably.  In fact, any solution of the new extended system (\ref{LN}) for which at least one the functional  ${\cal K}_{i}$'s takes non-zero values,  leads to a positive large function $\varepsilon_{i}$ and then the related rest energy would be larger that SSWPS rest energy, provided the constants  $A_{i}$'s  to be large numbers. We will show that there  is just a single non-trivial solution for which all ${\cal K}_{i}$'s would be zero simultaneously, i.e. the SSWPS (\ref{f2}). In fact, three conditions ${\cal K}_{i}=0$ ($i=1,2,3$) can be considered as three  non-linear PDE's as follows:
 \begin{eqnarray} \label{dfb}
 &&{\cal S}_{1}=\partial_{\mu}\theta\partial^{\mu}\theta-20=0,     \\&&
{\cal S}_{2}=\partial_{\mu}R\partial^{\mu}R-4R^{2}\ln(R)=0, \\&&
{\cal S}_{3}=\partial_{\mu}R\partial^{\mu}\theta=0.
\end{eqnarray}
Since the above equations  are three independent PDE's   for two fields $R$  and $\theta$, therefore, they may not be  satisfied simultaneously except for the SSWPS (\ref{f2}) and trivial vacuum state $R=0$ (for which just Eq.~(45) remains). So, for any arbitrary non-trivial solution, except the SSWPS, at least one of the functional ${\cal K}_{i}$'s (${\cal S}_{i}$'s) must  be a   non-zero function, and then if $A_{i}$'s  to be large numbers, at least one of the $\varepsilon_{i}$'s would be a non-zero positive large function which   lead to   rest energy   larger than the SSWPS rest energy. Accordingly,  we are sure that the rest energy of the SSWPS (\ref{f2}) is really a minimum among the other non-trivial solution, i.e. it is a soliton solution.

 To prove that the SSWPS (\ref{f2}) is really a stable object, we just considered
functions $\varepsilon_{i}$'s ($i = 1, 2, 3$) but we did not consider function $\varepsilon_{o}$. In the next section, it will be shown that the role of  the first part of the energy density $\varepsilon_{o}$, if $A_{i}$'s  to be large numbers, is physically  unimportant and  can be ignored   in the  stability considerations.


 In fact,  to  bring up an extended CNKG  model  (\ref{LN}) for which   the stability  of  the SSWPS (\ref{f2}) is  guaranteed appreciably in a simple straightforward conclusion, we select three special  linear combination of ${\cal S}_{i}$'s in Eqs.~(37), (38) and (39) for which   $\varepsilon_{i}$'s ($i=1,2,3$) would be definitely positive. In general,  it may be possible to choose other  combinations of ${\cal S}_{i}$'s for this goal. However, we intentionally introduced this special combination  (\ref{F})  as a good example of the extended CNKG systems (\ref{LN}) for better and simpler  conclusions.


\section{stability under small deformations}\label{sec5}

  In general, any arbitrary close solution or any small permissible deformed function of a non-moving  SSWPS (\ref{f2}) is introduced in the following forms:
\begin{equation} \label{so1}
R(x,y,z,t)=R_{s}(r)+\delta R(x,y,z,t) \quad \textrm{and} \quad \theta(x,y,z,t)=\theta_{s}+\delta \theta=\omega_{s}t+\delta \theta(x,y,z,t),
\end{equation}
where $\delta R$ and $\delta \theta$ (small variations) are    small functions  of space-time. Note that, the  permissible deformed functions $R(x,y,z,t)$ and $\theta(x,y,z,t)$ are considered to be  solutions of the new equations of motions (29) and (30) as well.
Now, if we insert  (\ref{so1}) in    $\varepsilon_{o}(x,y,z,t)$ and keep it to the first order of $\delta R$ and $\delta \theta$, then it yields
\begin{eqnarray} \label{so3}
&&\varepsilon_{o}(x,y,z,t)=\varepsilon_{os}(r)+\delta\varepsilon_{o}(x,y,z,t)= \left[\nabla R_{s} \cdot \nabla R_{s}+R_{s}^2\omega_{s}^2+V(R_{s})\right]+\nonumber\\&&
2\left[\nabla R_{s} \cdot \nabla(\delta R)+R_{s}(\delta R)\omega_{s}^2+
   R_{s}^{2}\omega_{s}(\delta\dot{\theta})+\frac{1}{2}\frac{dV(R_{s})}{dR_{s}}(\delta R)\right].
\end{eqnarray}
Note that, for a non-moving SSWPS, $\dot{R_{s}}=0$, $\nabla\theta_{s}=0$ and $\dot{\theta_{s}}=\omega_{s}=\sqrt{20}$. It is obvious that $\delta\varepsilon_{o}$ is not necessarily a positive definite function.

Now, let do this for the additional terms $\varepsilon_{i}$ ($i=1,2,3$). If we insert  a variation like (\ref{so1}) into  $\varepsilon_{i}$ ($i=1,2,3$), it yields
\begin{eqnarray} \label{so4}
&&\varepsilon_{i}(x,y,z,t)=\varepsilon_{is}+\delta\varepsilon_{i}=\delta\varepsilon_{i}=[3A_{i}(C_{is}+\delta C_{i})({\cal K}_{is}+\delta{\cal K}_{i})^{2}-A_{i}({\cal K}_{is}+\delta{\cal K}_{i})^{3}]=\nonumber\\&&
[3A_{i}(C_{is}+\delta C_{i})(\delta{\cal K}_{i})^{2}-A_{i}(\delta{\cal K}_{i})^{3}]\approx[3A_{i}C_{is}(\delta{\cal K}_{i})^{2}-A_{i}(\delta{\cal K}_{i})^{3}]\approx[3A_{i}C_{is}(\delta{\cal K}_{i})^{2}]>0
\end{eqnarray}
in which $\varepsilon_{is}=0$, ${\cal K}_{is}=0$ and $C_{is}$ referred  to the SSWPS and  $\delta{\cal K}_{i}$ and $\delta C_{i}$ are  in the same order of $\delta R$ and $\delta \theta$. Therefore, since $C_{i}>0$, according to Eq.~(\ref{so4}),
$\varepsilon_{i}$'s for small variations are always positive definite (as were  perviously  obtained  from Eqs.~(42), (43) and (44) generally). Note that $\delta{\cal K}_{i}$'s and $\delta\varepsilon_{o}$ are in the same order of magnitude of $\delta R$ and $\delta\theta$, but $\delta\varepsilon_{i}=\varepsilon_{i}$ ($i=1,2,3$) is proportional to  $A_{i}(\delta{\cal K}_{i})^{2}$.  The variation of total energy density is equal to $\delta\varepsilon=\delta\varepsilon_{o}+\sum_{i=1}^{3}\delta\varepsilon_{i}$. The stability is guaranteed if the variation of the total energy density  $\delta\varepsilon$ being positive for all possible arbitrary variations  in the modulus and phase functions (\ref{so1}). If one consider large values for constants $A_{i}$'s, this main goal confirms  effectively. Note that,  $\delta\varepsilon_{i}$'s are always positive but $\delta\varepsilon_{o}$ is not necessarily positive (c.f. \ref{so3} and \ref{so4}). If the order of $\delta\varepsilon_{o}$ for any arbitrary  small variation is greater than  $\delta\varepsilon_{i}$, it may be possible to see the decreasing behavior for the total rest energy $E_{o}$.
For example, consider $A_{i}=10^{40}$; therefore  the  order of magnitude of  variations $\delta R$ and $\delta\theta$, for which the SSWPS is not mathematically a stable object (i.e. the variations for which  $O(|\delta\varepsilon_{o}|)>O(\delta\varepsilon_{i})\approx A_{i}(\delta{\cal K}_{i})^{2}$ or $O(|\delta R|)+O(|\delta\theta|) >O(A_{i}[\delta R]^2)+O(A_{i}[\delta\theta]^2)+O(A_{i}|\delta R\delta\theta|)$),  is approximately less than $10^{-20}$, which is so small that can be ignored  in the stability considerations! For such so small variations, the total rest energy $E_{o}$ may be reduced  with a very small amounts equal  to the integration of $\delta\varepsilon_{o}$ over all the whole  space which again is a very small unimportant value. Therefore for large values of $A_{i}$'s, the  SSWPS  is effectively a stable object. In fact, this so small decreasing behaviour related to this fact that for a non-deformed (or for a very small deformed) SSWPS (\ref{f2}), the dominant    equations of motion are the  reduced  versions of the  equations of motion (29) and (30), i.e.   the same original CNKG equations  of motion (5) and (6). Note that, since  scalars ${\cal K}_{i}$'s  (or  ${\cal S}_{i}$) are three independent functions of $R$ and derivatives of $R$ and $\theta$, therefore, if constants $A_{i}$'s are large values, for any arbitrary small deformations, at least one of ${\cal K}_{i}$'s changes and takes non-zero values, which according to Eq.~(\ref{so4}), leads  to  the large increase in  the total rest energy. Although, the $A_{i}$'s parameters  take  very large values,   but they  won't affect the dynamical equations and the observable of the SSWPS (\ref{f2}). They just make it  stable and do not appear in any of the observable, i.e. they act like a stability catalyser.

\begin{figure}[ht!]
  \centering
  \includegraphics[width=120mm]{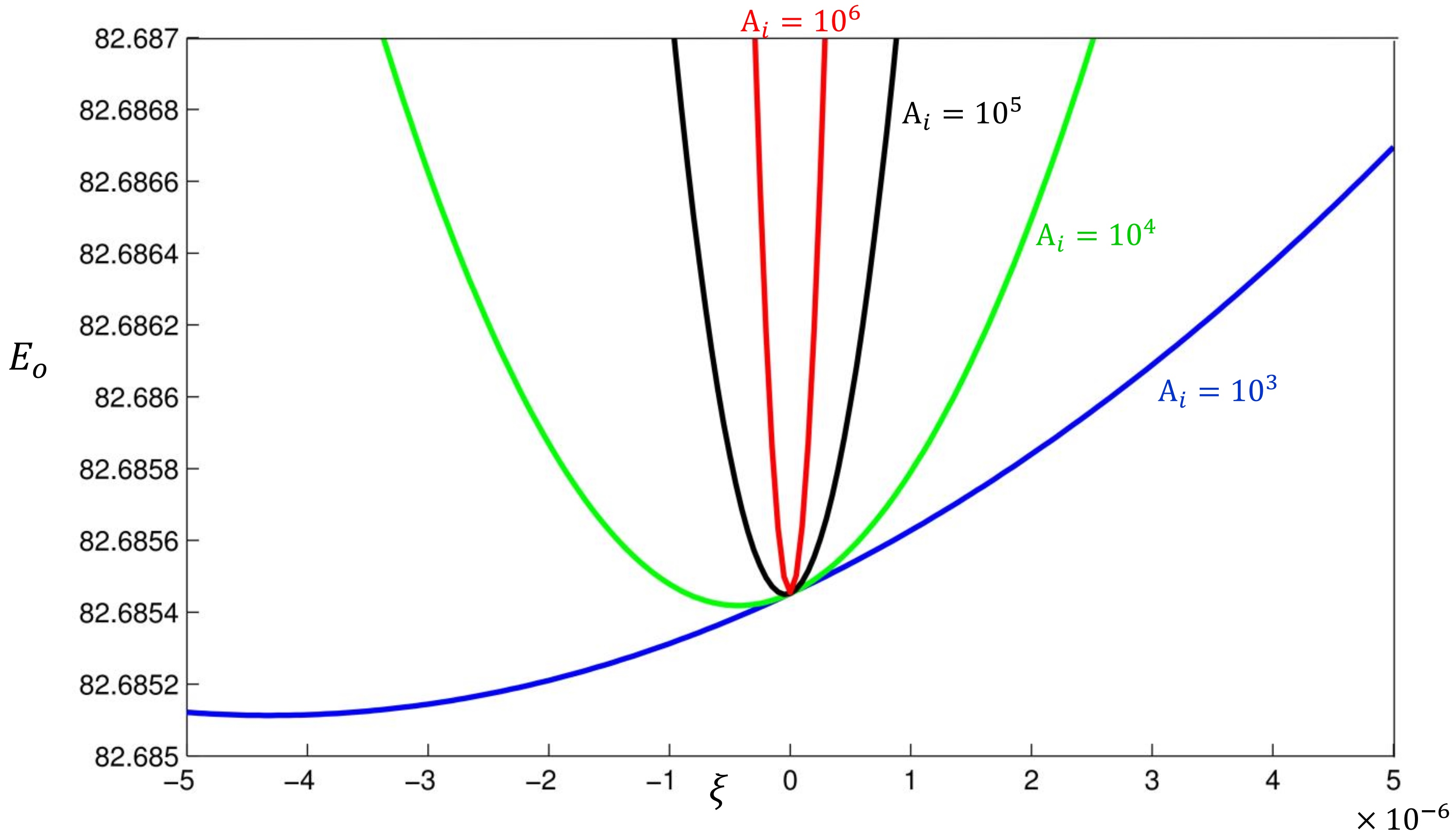}
  \caption{Variations  of  the total rest energy $E_{o}$ versus small $\xi$ for different $A_{i}$'s at $t=0$. We have a fixed  phase function and have considered  modulus  function changes according to relation (\ref{so6}).} \label{v1}
\end{figure}
\begin{figure}[ht!]
  \centering
  \includegraphics[width=120mm]{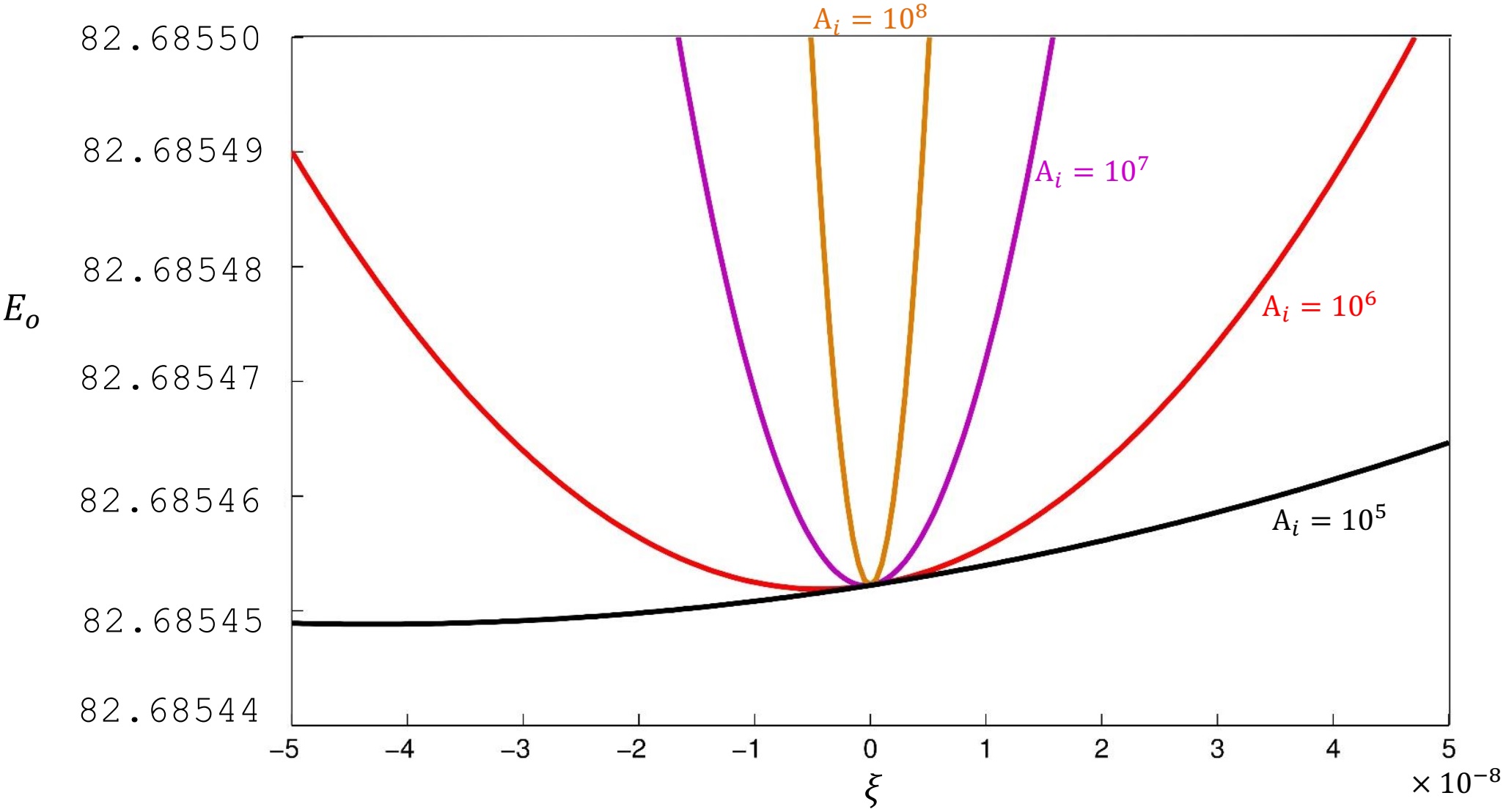}
  \caption{Variations  of  the total rest energy $E_{o}$ versus small $\xi$ for different $A_{i}$'s at $t=0$. We have a fixed  phase function and have considered  modulus  function changes according to relation (\ref{so6}).} \label{v2}
\end{figure}
\begin{figure}[ht!]
  \centering
  \includegraphics[width=120mm]{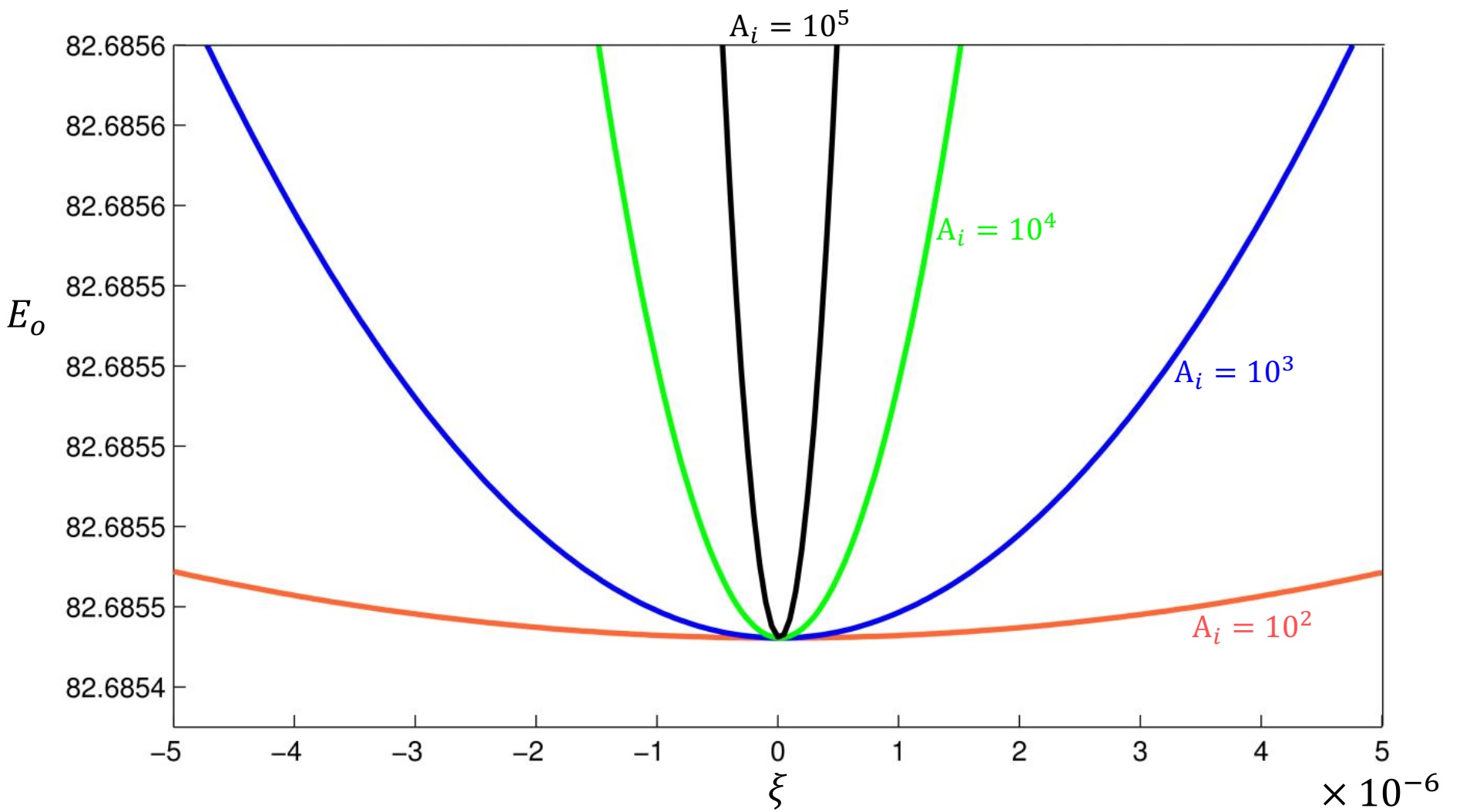}
  \caption{Variations  of  the total rest energy $E_{o}$ versus small $\xi$ for different $A_{i}$'s at $t=0$. We have a fixed modulus function   and have considered the phase function changes according to relation (\ref{var2}).} \label{v3}
\end{figure}
\begin{figure}[ht!]
   \centering
   \includegraphics[width=120mm]{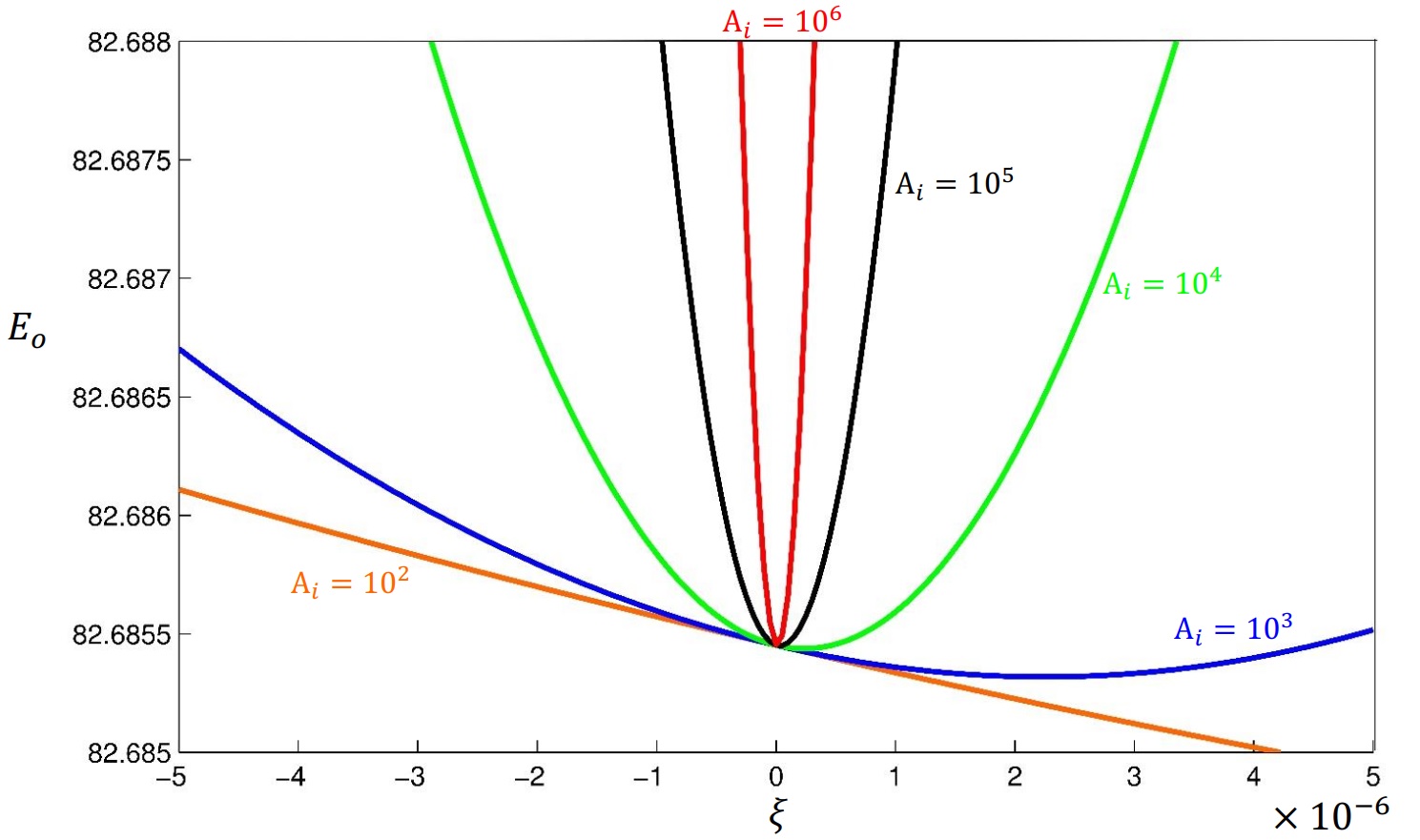}
   \caption{Variations  of  the total rest energy $E_{o}$ versus small $\xi$ for different $A_{i}$'s at $t=0$. We have fixed modulus function   and have considered the phase function changes according to relation (\ref{var2}).} \label{v4}
\end{figure}

In general, it is not possible for us to feel  the permissible deformations of the SSWPS for the new equations  of motion (29) and (30). Then let us  to consider an arbitrary  artificial (impermissible) deformation   at $t=0$ as follows:
\begin{equation} \label{so6}
R(x,y,z,t)= e^{- r^{2}}+\xi(1+t)e^{-r^2}\quad  \textrm{and} \quad \theta(x,y,z,t)=\theta_{s}=k_{\mu}x^{\mu},
\end{equation}
in which $\xi$ is  a small coefficient. Fig.~\ref{v1} and  Fig.~\ref{v2}, for this arbitrary variation (\ref{so6}), show that properly there is always a small range for which $E_{o}$ decreases  with very small values. Namely, in Fig.~\ref{v1}, for  $A_{i}=10^4$, there is not any minimum but if  we zoom on the curve  around the $\xi=0$, the output result  can be seen in Fig.~\ref{v2}, which shows that $\xi=0$ is not really a minimum. By  increasing $A_{i}$'s, this behavior never disappear, i.e.  there will be always a small range for which the arbitrary  variation (\ref{so6}) leads to a decreasing   behaviour for the total rest energy $E_{o}$. If we consider  extremely  large values  of $A_{i}$'s, this would  yield very small shift from $E_{o}(\xi=0)$, which is completely  unimportant and the  stability of the  solitary wave-packet solutions is enhanced appreciably. For more supported, we can introduce two  additional arbitrary impermissible  variations at $t=0$ as follows:
\begin{equation}\label{var2}
R(x,y,z,t)=e^{- r^{2}}\quad  \textrm{and} \quad \theta(x,y,z,t)=k_{\mu}x^{\mu}+\xi e^{-r^2},
 \end{equation}
and
\begin{equation}\label{var3}
R(x,y,z,t)=e^{-(1+\xi ) r^{2}}\quad  \textrm{and} \quad \theta(x,y,z,t)=k_{\mu}x^{\mu}.
\end{equation}
The expected results which obtained numerically for these  different  variations are shown in Fig.~\ref{v3} and Fig.~\ref{v4}.

Briefly, if one  considers large values for $A_{i}$'s, the SSWPS  is physically  a  stable object. The larger values of coefficients  $A_{i}$'s, similar to the pervious  Figs in this section,   leads to the greater  increase in  differences between the rest energy of the SSWPS and other close solutions.  In other words,  the larger values of $A_{i}$'s lead to  more stability.  As for a SSWPS, all ${\cal K}_{i}$'s are equal to zero  simultaneously, the related equations of motion  (29) and (30) are reduced  to the original forms (5) and (6) respectively, i.e. the dominant equations of motion are the same standard  CNKG equation (5) and (6) as well.
This means, if one asks  about the right equation of motion of the free SSWPS, our answer would be the same known CNKG Eq.~(\ref{eq}).
The role of the additional terms (${\cal K}_{i}$ dependent terms) which we consider in the new modified model (\ref{LN}), behave like a  strong force which fix  the SSWPS to a special form of modulus and phase function  (\ref{f2}), i.e. any deformation  in the  SSWPS arises to a strong force (that comes from the ${\cal K}_{i}$ dependent terms) and suppresses  the changes  and preserve the form of SSWPS. In other words, the nonstandard (${\cal K}_{i}$ dependent) terms for this modified model,   behave like a zero rest mass spook which surrounds the  particle-like solution and resist to any arbitrary  deformation. Therefore,  the ${\cal K}_{i}$ dependent terms, just guarantee the stability of the SSWPS and for a free non-deformed  version of that  are hidden.   As two SSWPS's approach each other to collide, the ${\cal K}_{i}$ dependent terms of these two SSWPS's become stronger and effectively  represent the interaction between them.
A SSWPS can move with different velocities and since it is a stable object we expect it to reappear in collisions, i.e. it is a   soliton.

\section{Summery and conclusion}\label{sec7}

Firstly, we reviewed some basic properties of the complex non-linear  Klein-Gordon (CNKG) equations in $3+1$ dimensions. Each CNKG equation may have some  non-dispersive solitary wave packet solutions  which can be identified by different rest frequencies ($\omega_{o}$).  For a moving solitary wave packet solution, the corresponding frequency is $\omega=\gamma\omega_{o}$ and it is proportional to the total energy, i.e. $E=\overline{h} \omega$ which $\overline{h}$ is just a Planck-like constant and is a function of the rest frequency. Moreover, it was found that a solitary  wave packet satisfies a deBroglie's-like wavelength-momentum relation i.e. $p =\overline{h}k$.

We had some arguments about  stability criterion. Briefly, a  stable solitary wave solution  is the one which  its rest energy is minimum among  the other close  solutions.   For a special CNKG system with Gaussian solitary wave-packet solutions (as a simple candidate of the CNKG systems),  it was shown that  there is not a stable solitary wave packet solution at all. Accordingly, to have a special stable solitary wave-packet solution (SSWPS) with a dominant standard CNKG equation of motion, we have to add three special terms to the original  CNKG Lagrangian density. It was shown that for the new extended CNKG system (i.e. the original CNKG Lagrangian density $+$ three proper additional terms), the stability  of the SSWPS satisfied appreciably. In fact, these additional terms behave like a zero rest mass spook  which surrounds the SSWPS and resist to any arbitrary deformations. For the new extended CNKG system, in general,  there are complicated equations of motion, which just for the SSWPS, are reduced  to the original versions. In other words, for the new extended system, the dominant equation of motion for the SSWPS would be the  same original CNKG versions as we expected. For this modified model, there are some  free parameters $A_{i}$'s ($i=1,2,3$)  which larger values of that imposes stronger  stability on the SSWPS, i.e. the difference   between the rest energy of the SSWPS and the rest energies of the other close
solutions, increases with increasing the amount of constant $A_{i}$’s.

\end{document}